\documentclass[aps,prl,reprint,superscriptaddress,longbibliography,showpacs,nobibnotes,nofootinbib]{revtex4-2}

\usepackage[utf8]{inputenc}
\usepackage[T1]{fontenc}
\usepackage[]{lmodern}
\usepackage[]{microtype}

\usepackage[american]{babel}
\usepackage[]{csquotes}

\usepackage[]{amsmath}
\usepackage[]{amssymb,bbold}
\usepackage[mathscr]{euscript}
\usepackage{siunitx}

\newcommand{\bra}[1]{\ensuremath{%
    {\mkern -3mu}\left\langle#1\right\rvert{\mkern -3mu}%
}}
\newcommand{\ket}[1]{\ensuremath{%
    {\mkern -3mu}\left\lvert#1\right\rangle{\mkern -3mu}%
}}
\newcommand{\mel}[3]{\ensuremath{\bra{#1}#2\ket{#3}}}
\newcommand{\braket}[2]{\ensuremath{%
    {\mkern -3mu}\left\langle#1%
    \right\rvert
    {\mkern -11mu}%
    \left\lvert%
    #2\right\rangle{\mkern -3mu}%
}}
\newcommand{\ketbra}[2]{\ensuremath{%
    \ket{#1}%
    {\mkern -3mu}%
    \bra{#2}%
}}

\usepackage{siunitx}
\AtBeginDocument{\RenewCommandCopy\qty\SI}
\DeclareSIUnit\hopping{\mathit t_0}

\usepackage[]{graphicx}
\usepackage[percent]{overpic}
\usepackage{tikz-cd}
\usepackage[export]{adjustbox}
\usepackage[figurename=Figure]{caption}
\usepackage[]{subcaption}

\usepackage[subpreambles=true]{standalone}
\usepackage{comment}

\usepackage{ragged2e}
\DeclareCaptionJustification{plain}{\justifying}
\usepackage[]{hyperref}
\hypersetup{
    colorlinks=true,
    urlcolor=blue,
    citecolor=blue,
    linkcolor=blue
}
\usepackage[capitalize]{cleveref}

\usepackage{xcolor}
\definecolor{nodetuning}{rgb}{0.403, 0.678, 0.325}
\definecolor{interactdetuning}{rgb}{0.824, 0.208, 0.169}
\definecolor{localdetuning}{rgb}{0.949, 0.663, 0.231}
\definecolor{fillingdetuning}{rgb}{0, 0, 0.961}
\definecolor{greedydetuning}{rgb}{0.459, 0.078, 0.4902}

\DeclareMathOperator{\Tr}{Tr}

\DeclareMathOperator{\hc}{h.c.}

\usepackage{titlesec}

\titlespacing{\paragraph}{1em}{0em}{0.5em}
\let\originalparagraph\paragraph
\renewcommand{\paragraph}[2][.---]{\originalparagraph{#2#1}}

\usepackage{orcidlink}

\newcommand{\qmoperator}[1]{\ensuremath{\hat{#1}}}
 \newtheorem{theorem}{Theorem}

\usepackage{comment}

\begin{document}
	
\title{\texorpdfstring{Identifying slow relaxation in many-body quantum systems through\\state-graph geometry and state-graph heterogeneity}
{Identifying slow relaxation in many-body quantum systems through state-graph geometry and state-graph geometry}}

\author{Heiko Georg Menzler\,\orcidlink{0009-0001-9530-7392}}
\email{heiko.menzler@uni-goettingen.de}
\affiliation{Institut für Theoretische Physik, Georg-August-Universität Göttingen, D-37077 Göttingen, Germany}

\author{Tom Ben-Ami\,\orcidlink{0000-0001-7513-6701}}
\email{tom.ben.ami@uni-a.de}
\affiliation{Theoretical Physics III, Center for Electronic Correlations and Magnetism, Institute of Physics, University of Augsburg, D-86135 Augsburg, Germany}
\affiliation{Max-Planck-Institut für Physik komplexer Systeme, Nöthnitzer Straße 38, Dresden 01187, Germany}

\begin{abstract}
    We adapt tools from the theory of quantum random walks to investigate slow relaxation dynamics through the many-body state graph.
    Specifically, we construct a probe of heterogeneity between basis states defined using hitting times derived from the unitary time-evolution operator.
    We find that the state-graph geometry, encoded by the pairwise hitting time of basis states, is a highly sensitive indicator of slow relaxation dynamics in a variety of systems.
    We study three paradigmatic models: the Rosenzweig-Porter model, the quantum East model, and the triangular lattice gas model, exhibiting a sudden onset of slow dynamics upon tuning of a control parameter.
    As a global characterization of the graph geometry, we analyze the spectral radius of the hitting matrix. 
    We find that it increases sharply at the onset of slow dynamics, spanning many orders of magnitude, with a characteristic crossing point at the transition.  
    Our work provides a geometric framework for describing and identifying phases with slow relaxation using a unified graph-theoretic formalism. 
\end{abstract}

\date{\today}

\maketitle

\paragraph{Introduction}
Graph-theoretic descriptions arise naturally in the quantum dynamics of systems with discrete Hilbert spaces~\cite{farhi1998quantum,kempe2005discrete,venegas2012quantumw}, as exemplified by continuous-time quantum walks (CTQWs). 
Such CTQWs generalize classical Markov chains~\cite{aharonov2001quantumwalks,mlken2011continuous} through the unitary time-evolution generated by the Schrödinger equation. Consequently, they provide a natural framework for studying quantum dynamics on abstract graphs.

In quantum many-body systems such graph structures emerge naturally: systems with discrete Hilbert spaces, such as spin and lattice systems, can be mapped onto a single-particle hopping problem on a graph. Thus, graph-theoretic methods developed for CTQWs can be used to study quantum many-body dynamics of closed systems. 

Generically, isolated quantum systems are expected to thermalize under unitary dynamics~\cite{rigol2008thermalization}, as captured by the \emph{Eigenstate Thermalization Hypothesis} (ETH)~\cite{deutsch1991quantum, srednicki1994chaos, dallesio2016quantum, deutsch2018eigenstate}.
However, ETH can fail due to disorder effects, kinetic and energetic constraints, integrability, and related mechanisms~\cite{basko2006metal, sala2020ergodicity, yang2020hilbert, abanin2019colloquium}. 
Beyond outright ETH violation, some systems exhibit anomalously slow relaxation dynamics in certain parameter regimes, due to kinetic constraints and bottlenecks in Hilbert space~\cite{lan2018quantum, pancotti2020quantum, brighi2023hilbert, rakovszky2024bottlenecks}.
Recent approaches based on the many-body state graph formalism have emerged as powerful tools for characterizing slow dynamics and ETH violations in multiple systems~\cite{menzler2025graph,ben2025many, tan2025interference, nicolau2025fragmentation, jonay2025localized,menzler2026graph}. However, a dynamical notion of geometry that captures how the Hamiltonian explores this state graph remains unexplored.

\begin{figure}[!b]
    \centering
    \begin{overpic}{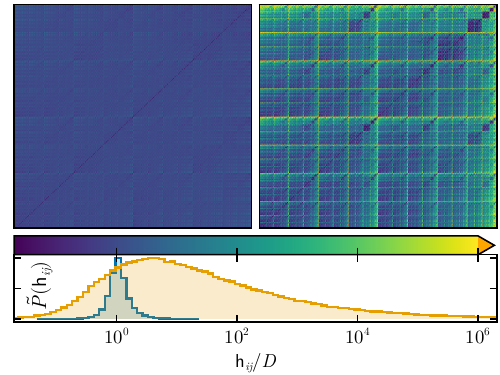}
        \put(5,68){\color{white}Quantum East}   
        \put(5,34){\color{white}(a) $s=-0.2$}   
        \put(55,34){\color{white}(b) $s=1$}   
        \put(90,19){(c)}   
    \end{overpic}
    {\phantomsubcaption\label{fig:distmat_qe_s=-0.2}}
    {\phantomsubcaption\label{fig:distmat_qe_s=1}}
    {\phantomsubcaption\label{fig:distmat_qe_distshape}}
    \vspace{-1.65em}
    \caption{
        Hitting matrix $\qmoperator{\mathsf{h}}$ in the QEM with $L=10$ ($D=2^{10}$) in the \subref{fig:distmat_qe_s=-0.2} fast-dynamics ($s = -0.2$) and \subref{fig:distmat_qe_s=1} slow dynamics ($s=1$) regime.
        Large values beyond the colorbar-cutoff up to $\mathsf{h}_{ij}/D \approx 10^{15}$ are shown in orange.
        \subref{fig:distmat_qe_distshape} Normalized distribution $\tilde{P}(\mathsf{h}_{ij})$ of matrix elements for \subref{fig:distmat_qe_s=-0.2} (blue) and \subref{fig:distmat_qe_s=1} (orange) such that the maximum of both distribution is the same.
    }
    \label{fig:introfigure_dists}
\end{figure}

Building on this connection between CTQWs on graphs and quantum many-body systems, we construct a Hamiltonian-induced geometry on the many-body state graph for characterizing thermalization and heterogeneous relaxation dynamics. 
We use the hitting time, defined as the time of traversal between basis states~\cite{redner2001guide,aldous2014fill}, as a probe of the state-graph's geometry.
The resulting probe encodes the structure of the unitary dynamics induced by the system's Hamiltonian.
We apply this construction on systems exhibiting a broad distribution of time scales, i.e., slow dynamics.
The distribution of hitting times reveals persistent heterogeneity on the state graph up to extremely late times, characterizing the underlying time-scale distributions.
Furthermore, structural measures on the resulting hitting matrix provide insight into the sudden onset of state-graph heterogeneity.
Our results suggest that state-graph heterogeneity sensitively characterizes slow dynamics.

Using this formalism, we show that the sudden onset of slow dynamics is characterized by a sharp change in the geometry of the state graph in three paradigmatic models: the Rosenzweig-Porter model (RPM)~\cite{rosenzweig1960repulsion,kravtsov2015random}, the quantum East model (QEM)~\cite{vanhorssen2015dynamics,pancotti2020quantum} and the triangular lattice gas model (TLGM)~\cite{lan2018quantum,royen2024enhanced}.
While the microscopic mechanisms underlying slow dynamics vary across models, the geometric signatures we find are universal. 
Across all models, the spectral radius of the Hamiltonian-induced hitting matrix sensitively distinguishes homogeneous and heterogeneous state-graph geometries, corresponding to fast and slow-dynamics regimes respectively. 
This is illustrated by the hitting matrices for the QEM in \cref{fig:introfigure_dists}.
Moreover, our construction reveals that structural probes of geometry on the hitting matrix, like the spectral radius, induces crossing points that indicate changes in the scaling behavior at infinite temperature. 

\paragraph{State Graph Geometry}
Throughout this work, we study the state graphs of many-body quantum systems. 
The nodes of the state graph represent many-body basis states $\ket{i}$, here chosen from the Fock basis, while edges are determined by the representation of the Hamiltonian in the same basis. 
Non-zero off-diagonal matrix elements correspond to edges between nodes, whereas diagonal matrix elements correspond to self-edges.

On a graph, classical notions of distance can be defined through random walk dynamics, such as diffusion-based distances or resistance distances~\cite{doyle1984random,klein1993resistance,lovasz1993random,chandra1996theelectrical,coifman2006diffusionmaps}.
We define the kernel function as ${k_{ij}(T) = \int_0^T P(i \to j; t) \,\mathrm{d}t}$, where ${P(i\to j; t)}$ is the probability that a random walker starting from node $i$ is found at node $j$ after time $t$.
We extend this construction to many-body quantum systems and their dynamics on the many-body state graph, whose nodes correspond to basis states $\{\ket{i}\}$.
The dynamics are generated through the unitary evolution $\qmoperator U(t) = e^{-i \qmoperator H t}$. Born's rule gives the transition probability between basis states $\ket{i}$ and $\ket{j}$ after time $T$ as
\begin{align}
    k_{ij}(T) = \int_{0}^{T} | \mel{i}{\qmoperator U(t)}{j}|^2\,\mathrm{d}t\,.
    \label{eq:kernel_continuous}
\end{align}
The kernel function, which quantifies the similarity of $\ket{i}$ and $\ket{j}$, is symmetric and monotonically increasing.
Up to a factor of $T^{-1}$, $\qmoperator k$ is also otherwise known as the average mixing matrix~\cite{coutinho2024selected}.
We define the hitting time~\cite{redner2001guide,aldous2014fill} between two nodes $\ket{i}$, $\ket{j}$ as
\begin{align}
    \mathsf{h}_{ij} = \min\{T: k_{ij}(T) \geq C\}\,,
    \label{eq:distance_generic}
\end{align}
providing a notion of distance on the state graph~\cite{chandra1996theelectrical}.
$\mathsf{h}_{ij}$ is the earliest time at which the cumulative transition probability from node $i$ to node $j$ reaches the prescribed threshold $C$.
$C$ acts as a scale parameter, resolving different features on the graph depending on its choice~\cite{coifman2006diffusionmaps}.
Throughout this work, we choose $C$ to be large but finite so that the late-time behavior of $k_{ij}(T)$ dominates.
Similar quantum hitting time constructions have been studied before in single-particle quantum walks on graphs~\cite{farhi1998quantum,childs2002anexample,childs2004spatial,szegedy2004quantum,mlken2011continuous,magniez2012hitting} and on hypercubes~\cite{moore2002quantumwalks,shenvi2003quantumrandom}.

Hitting times are closely related to notions of distance and geometry on graphs. 
In settings where $\mathsf h_{ii} = 0$ arises naturally, such as for discrete quantum walks (see End Matter), the hitting matrix $\qmoperator{\mathsf{h}}$ can be directly interpreted as a distance matrix on the state graph.
This notion of geometry carries over to the more general case $\mathsf h_{ii}\neq 0$.
While the kernel matrix $\qmoperator k$ offers a notion of transfer rate/transfer velocity, the hitting matrix $\qmoperator{\mathsf{h}}$ offers the inverse interpretation of resistance/distance for transferring probability weight from $\ket{i}$ to $\ket{j}$.
Alternatively, kernel-based distance functions can be constructed using diffusion maps~\cite{coifman2006diffusionmaps}; exploring such diffusion maps in quantum systems is left for future work.
Our construction connects graph-theory developed in the context of quantum walks to the emergent geometry of state graphs of interacting many-body systems. 

\paragraph{Models}
Throughout our analysis, we study three models, which represent three distinct mechanisms for slow dynamics.
First, the Rosenzweig-Porter model (RPM) is a random-matrix ensemble described by ${\qmoperator H_\mathrm{RPM} = \qmoperator H_0 + D^{-\gamma/2}}\qmoperator H_\mathrm{GOE}$, where $\qmoperator H_0$ is a diagonal matrix with random normal entries, and $\qmoperator H_\mathrm{GOE}$ is a random matrix from the Gaussian orthogonal ensemble.
The Hamiltonian $\qmoperator H$ has dimension $D=2^L$ and a tunable system parameter $\gamma$.
The RPM exhibits three established regimes~\cite{kravtsov2015random}; ergodic extended at $\gamma < 1$, non-ergodic extended at $1 < \gamma < 2$, and a localized regime at $\gamma > 2$.
Leaving the ergodic regime, the model exhibits slow relaxation due to fractal eigenstate structures.

Second, the quantum East model (QEM) is a kinetically constrained model described by the Hamiltonian ${\qmoperator H_\mathrm{QEM} = \frac{1}{2}\sum_{\ell=1}^{L-1} \qmoperator n_\ell (e^{-s} \qmoperator \sigma^x_{\ell+1} - \mathbb{1})}$.
In the QEM, $s$ drives the system into a regime of slow dynamics (more details, e.g., boundary conditions, are discussed in~\cite{SM}).
For ${s < 0}$ the system exhibits fast dynamics where autocorrelations decay quickly to their ensemble value, while for ${s \gtrsim 0}$ autocorrelations remain finite on small systems~\cite{pancotti2020quantum,menzler2025graph}.
Slow dynamics in the QEM is characterized by a qualitative change in the autocorrelation function decay.
The accurate position of the point for the onset of slow dynamics at high temperature is a subject of debate in the literature. 

Third, the triangular lattice gas model (TLGM) is an example of a kinetically constrained model on a triangular ladder lattice geometry.
Its Hamiltonian is given 
by~\cite{lan2018quantum} $\qmoperator H_\mathrm{TLG} = \sum_{\langle \ell,\ell^\prime \rangle} \qmoperator C_{\ell \ell^\prime} \big[ 
        -\unit{\hopping}(\qmoperator a_\ell^\dagger \qmoperator a_{\ell^\prime} + \hc)
        + V (\qmoperator n_\ell (\mathbb{1} - \qmoperator n_{\ell^\prime}) + (\mathbb{1} - \qmoperator n_\ell) \qmoperator n_{\ell^\prime})
    \big]
$,
where $\langle \ell, \ell^\prime\rangle$ denotes pairs of nearest neighbor sites on a triangular ladder and $\qmoperator C_{\ell\ell^\prime} = \mathbb{1} - \prod_{k \in \mathcal{N}(\ell,\ell^\prime)} \qmoperator n_k$ with $\mathcal{N}(\ell,\ell^\prime)$ denoting the joint neighborhood of $\ell$ and $\ell^\prime$ on the triangular ladder.
Here, the ratio $V/t_0$ tunes the system into a slow-dynamics regime.
In contrast to the QEM, the TLGM features slow dynamics in the form of metastable plateaus~---~autocorrelations decay to their ensemble average, but they do so only at late times~\cite{lan2018quantum}.
The metastable behavior of the TLGM can be understood via perturbation theory: for $V/t_0\gtrsim 1$, plateaus are generated by high-order perturbation theory processes~\cite{royen2024enhanced}.
The occurrence of these metastable plateaus characterizes the onset of slow dynamics and is accompanied by dynamical heterogeneity in real space~\cite{lan2018quantum}.
As with the QEM, the point for the onset of slow dynamics has not yet been accurately determined.

\begin{figure}[t]
    \centering
    \includegraphics[]{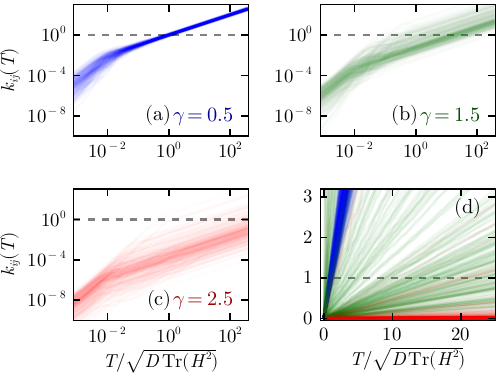}
    {\phantomsubcaption\label{fig:kernel_rpm_c=0.5}}
    {\phantomsubcaption\label{fig:kernel_rpm_c=1.5}}
    {\phantomsubcaption\label{fig:kernel_rpm_c=2.5}}
    {\phantomsubcaption\label{fig:kernel_rpm_linear}}
    \caption{
        Kernel function $k_{ij}(T)$ for a representative state $\ket{i}$ and other states $\ket{j}$ in the computational basis in the RPM for \subref{fig:kernel_rpm_c=0.5} ergodic extended phase $\gamma=0.5$, \subref{fig:kernel_rpm_c=1.5} non-ergodic extended phase $\gamma=1.5$ and \subref{fig:kernel_rpm_c=2.5} localized phase $\gamma=2.5$.
        \subref{fig:kernel_rpm_linear} combines data from the three other subplots, showing linear behavior of $k_{ij}(T)$ at large $T$. 
        Horizontal dashed lines indicate a cutoff value $C=1$.
    }
    \label{fig:kernel_rpm}
\end{figure}

\paragraph{Emergent Geometry of Many-body State Graphs}
Using the formalism above, we now investigate the emergent geometry induced by quantum hitting times on the many-body state graph. We find that at late times the kernel depends only on overlaps between basis states and eigenstates, relating the emergent geometry of the state graph to the eigenstate structure of the system. 

Expanding \cref{eq:kernel_continuous} in the eigenbasis of $\qmoperator H$, we can solve the integral and obtain the asympotic limit at large $T$,
\begin{align}
    \lim_{T\to\infty} k_{ij}(T)/ T = \sum_{\nu} |\braket{i}{\nu}|^2 | \braket{j}{\nu}|^2\,,
    \label{eq:kernel_continuous_limit}
\end{align}
assuming a system without degeneracies.
See more details in~\cite{SM}.
We thus denote the asymptotic slope $\alpha_{ij}$, such that $k_{ij}(T) \sim \alpha_{ij}T$ for large $T$.
Consequently, in this limit and by plugging into \cref{eq:distance_generic}, we obtain $\mathsf{h}_{ij} = \alpha_{ij}^{-1} C $, where $\alpha_{ij}^{-1}$ fully characterizes the long-time behavior of $\mathsf{h}_{ij}$.

The emergence of the linear regime is clearly visible in the RPM, which we use as a representative example. 
\cref{fig:kernel_rpm} shows $k_{ij}(T)$ as a function of $T$ for the RPM, in its different dynamical regimes. We fix a single node $i$, and see how $k_{ij}$ varies for all basis states. The spread of $k_{ij}$ reflects the wide range of values of quantum hitting times between $i$ and the rest of the nodes. 
Across all regimes we observe non-universal behavior at early  times followed by saturation to a linear ramp.
In the ergodic regime $\gamma=0.5$, \cref{fig:kernel_rpm_c=0.5}, $k_{ij}$ exhibit a narrow spread for all $T$ within the time frame of the linear ramp. 
In contrast, for $\gamma=1.5$, \cref{fig:kernel_rpm_c=1.5}, and $\gamma=2.5$, \cref{fig:kernel_rpm_c=2.5}, show a broader spread of $k_{ij}$. 
In \cref{fig:kernel_rpm_linear} we compare the linear ramp across the three parameter points.

In the following, we focus exclusively on the long-time regime and normalize $\mathsf{h}_{ij}/C=\alpha_{ij}^{-1}$, since at the late time regime of linear growth of all $k_{ij}$, $C$ carries no physical information.
Additionally, in \cref{fig:kernel_rpm} we observe that choosing $C=1$ is often sufficient for probing the late time regime of $k_{ij}$ around the Heisenberg time $t_H \sim \sqrt{D\Tr(H^2)}$.

\begin{figure*}[!t]
    \centering
    \hspace{1pt}\begin{overpic}{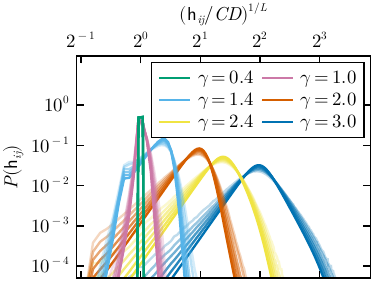}
        \put(25, 52){(a)}
    \end{overpic}
    \begin{overpic}{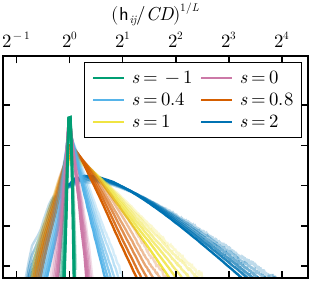}
        \put(10, 62){(b)}
    \end{overpic}
    \begin{overpic}{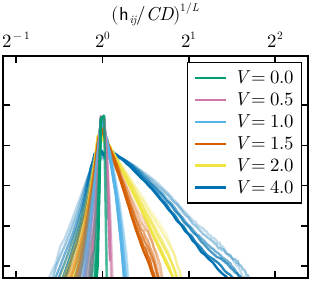}
        \put(10, 62){(c)}
    \end{overpic}
    \vspace{0.2em}
    \begin{overpic}{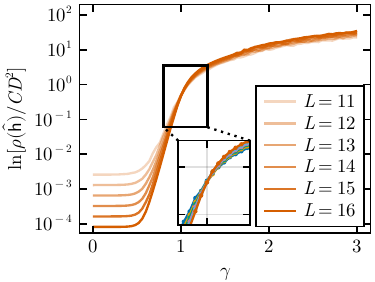}
        \put(25, 66.5){(d) Rosenzweig-Porter}
    \end{overpic}
    \begin{overpic}{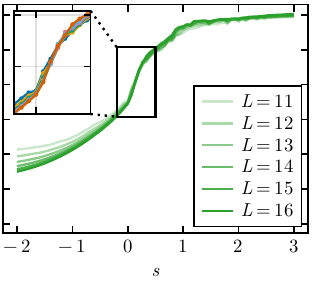}
        \put(6, 23){(e) Quantum East}
    \end{overpic}
    \begin{overpic}{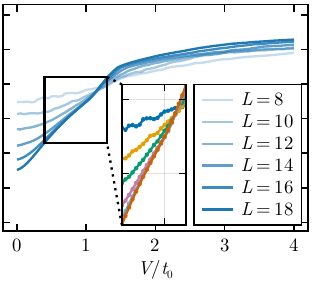}
        \put(6, 80){(f) Triangular lattice gas}
    \end{overpic}
    
    {\phantomsubcaption\label{fig:rpm_distance_dist}}
    {\phantomsubcaption\label{fig:qe_distance_dist}}
    {\phantomsubcaption\label{fig:tlg_distance_dist}}
    {\phantomsubcaption\label{fig:rpm_perron}}
    {\phantomsubcaption\label{fig:qe_perron}}
    {\phantomsubcaption\label{fig:tlg_perron}}
    
    \caption{
        Characterization of heterogeneous relaxation through $\qmoperator{\mathsf{h}}$.
        \subref{fig:rpm_distance_dist}-\subref{fig:tlg_distance_dist}~Distribution of the hitting matrix elements $\mathsf{h}_{ij}$ across system parameters in the \subref{fig:rpm_distance_dist}~RPM $L=\SIrange[]{10}{15}{}$, \subref{fig:qe_distance_dist}~QEM model $L=\SIrange{10}{15}{}$ and \subref{fig:tlg_distance_dist}~the TLGM $L=\SIrange{14}{19}{}$ with $N = \lfloor L/2\rfloor$, respectively.
        Lighter colors indicate smaller system sizes.
        The rescaling $\mathsf{h}_{ij}^{1/L}$ ensures that distributions have a comparable support across the investigated system sizes.
        \subref{fig:rpm_perron}-\subref{fig:tlg_perron}~Spectral radius (largest eigenvalue) of the hitting matrix $\rho(\qmoperator{\mathsf{h}})$ against system parameters in the \subref{fig:rpm_perron}~RPM, \subref{fig:qe_perron}~QEM and \subref{fig:tlg_perron}~TLGM for various system sizes.
        Insets show a zoom-in of the crossings.
        The logarithm of $\rho(\qmoperator{\mathsf{h}})$ is shown on a log-scale, making the visualization effectively $\log_{10}[\ln[\ldots]]$.
        For the RPM single realizations are shown, with every different $\gamma$ or $L$ being independent.
    }
    \label{fig:distmat_main_figure}
\end{figure*}

\paragraph{Geometric Signatures of state-graph heterogeneity}
We now show that heterogeneous relaxation in our three models manifests itself as a restructuring of the geometry encoded in the quantum hitting matrix, which is visible both in the distribution of $\mathsf{h}_{ij}$ and in its global spectral properties.

The distributions of $\mathsf{h}_{ij}$ [\cref{fig:rpm_distance_dist,fig:qe_distance_dist,fig:tlg_distance_dist}] capture the onset of slow dynamics across all three models.
In the fast dynamics regime, the distributions are sharply peaked at $\mathsf{h}_{ij}/C=D$, indicating homogeneous state-graph geometry.
This follows from substituting random-matrix-like states, ${|\braket{i}{\nu}|^2\sim 1/D \, \forall i, \nu}$, into \cref{eq:kernel_continuous_limit}.
In contrast, in the slow-dynamics regime the distributions deviate strongly from the random eigenstates assumption, becoming significantly broader. 
This indicates heterogeneous dynamical behavior across basis states.
We rescale the hitting matrix as $(\mathsf{h}_{ij}/CD)^{1/L}$ to prevent an exponential growth of the distributions support.
This is demonstrated via a finite-size analysis, indicated by the color shading.
Across different system sizes, the distributions only show small deviations after rescaling, indicating that distributions of $\mathsf{h}_{ij}$ exhibit exponential system-size sensitivity to the onset of slow dynamics.

In addition to the distributions, $\qmoperator{\mathsf{h}}$ undergoes a structural change when crossing into the slow-dynamics regime. 
We probe this through the spectral radius $\rho(\qmoperator{\mathsf{h}})$, i.e., the largest eigenvalue of $\qmoperator{\mathsf{h}}$.
\Cref{fig:rpm_perron,fig:qe_perron,fig:tlg_perron} show $\rho(\qmoperator{\mathsf{h}})$ as a function of the system parameter across our three models.
In the fast dynamics regime, the behavior of $\rho(\qmoperator{\mathsf{h}})$ follows from the random eigenstate assumption.
In this limit, $x = (1, 1, \ldots, 1)^T$ is an eigenvector, with eigenvalue $D^2$, of $\qmoperator{\mathsf{h}}$. 
Following the Perron-Frobenius theorem~\cite{meyer2000matrix}, this eigenvalue is the spectral radius of $\qmoperator{\mathsf{h}}$ and the homogeneous structure of $x$ reflects the homogeneity of $\qmoperator{\mathsf{h}}$. 
Therefore, we rescale the spectral radius with $D^2$.

We find that $\rho(\qmoperator{\mathsf{h}})$ is highly sensitive to the onset of slow dynamics across all three models.
When increasing system size, $\rho(\qmoperator{\mathsf{h}})/CD^2$ approaches unity from above in the fast dynamics regime.
At the onset of slow dynamics, we observe a crossing point across all three models. 
In the slow dynamics regime, $\rho(\qmoperator{\mathsf{h}})/CD^2$ is large and increases with system size, reflecting the breakdown of the homogeneous state-graph geometry.
The most drastic example is given by the QEM at $L=12$, where $\rho(\qmoperator{\mathsf{h}})/CD^2$ changes from approximately~$1.01$ at $s=-2$, to~$>10^{40}$ at $s=2$.
This highlights the extreme sensitivity of our measure to the onset of the slow dynamics.

\paragraph{Discussion}
In this work, we have shown how quantum many-body systems naturally induce an emergent geometry when expressed in a chosen (e.g., Fock) basis. 
An important distinction to previous approaches is that the geometric probe introduced here is analogous to a resistance-distance, i.e., all dynamical paths on the graph contribute. In contrast, shortest-path distances are determined only by the single shortest path connecting two nodes.
Consequently, shortest-path distances neglect contributions of additional dynamical pathways or quantum mechanical interference effects, and do not fully capture the global structure of the state graph.
Unlike shortest-path graph distances, which are based solely on connectivity and are usually unitless, quantum hitting times are formulated as dynamical quantities and thus carry units of time. 
Since quantum hitting times are defined through the unitary evolution $\qmoperator U$, the resulting geometry is naturally tied to the characteristic energy scale and units of the system.

While the quantum hitting matrix may appear abstract in the context of many-body systems, it is directly connected to the decay of autocorrelation functions of projectors in the state-graph basis.  
Consider the projector $\qmoperator n = \sum_{i \in S} \ketbra{i}{i}$ where $S = \{i : \qmoperator n \ket{i} = \ket{i}\}$, which includes both local and non-local projectors.
The corresponding time-averaged autocorrelation function is
\begin{align}
\overline c(T) = \frac{1}{DT}\int_0^T \Tr(\qmoperator n(t)\qmoperator n)\,\mathrm{d}t = \frac{1}{DT}\sum_{(i, j)\in S^2} k_{ij}(T)\,,
\end{align}
with $\lim_{T\to\infty}\overline c(T) = \frac{1}{D^2}\sum_{(i, j)\in S^2} (\mathsf h_{ij}/CD)^{-1}$.
A narrow distribution of $\mathsf{h}_{ij}$ implies that basis states contribute comparably to the autocorrelation function, leading to rapid and homogeneous relaxation to its infinite-temperature value. 
In contrast, broad distributions of $\mathsf{h}_{ij}$ imply heterogeneous relaxation dynamics and can give rise to long-lived memory effects or non-decaying autocorrelations (see~\cite{SM} for a detailed discussion).

Finally, our findings on the spectral radius point to universal behavior in the fast-dynamics regime across all three models. 
In the thermal fast-dynamics regime, $\rho(\qmoperator{\mathsf{h}})$ approaches its universal minimum corresponding to a class of all-to-all connected graphs (see~\cite{SM} for proof). 
In contrast, in the slow-dynamics regime $\rho(\qmoperator{\mathsf{h}})$ does not exhibit a single universal scaling or structure across the three models we study. 
The analysis of $\rho(\qmoperator{\mathsf{h}})$ unifies the different slow dynamics phenomena as a breakdown of homogeneity on the state-graph and reinterprets the onset of slow dynamics as a structural change in the geometry of the underlying graph that governs relaxation dynamics of selected autocorrelation functions.

\paragraph{Outlook}
Our framework provides a potential experimental probe for eigenstate structure through dynamical correlation measurements. In its original form, measuring $k_{ij}(T)$ requires evolving the state $\ket{i}$ in time and obtaining full measurement statistics of the probability of detecting state $\ket{j}$ at time $T$. In quantum many-body systems that exhibit exponentially large Hilbert spaces this is experimentally unfeasible.

However, this difficulty can be circumvented by reformulating the problem in terms of operator correlations. 
Writing $|\braket{i}{j(t)}|^2 = \braket{i}{j(t)}\braket{j(t)}{i} = \Tr(\qmoperator{O}_i \qmoperator{O}_j(t))$, with $\qmoperator{O}_k = \ketbra{k}{k}$,
the reformulation naturally extends from states to operators, making it directly accessible to experimentally relevant observables.
In particular, one may consider local or coarse-grained operators such as particle-number projectors on subsystems.
This is closely related to the recently experimentally studied subsystem Loschmidt echos~\cite{karch2025probing}. 
However, while the Loschmidt echo typically focuses on return probabilities, here our framework resolves transition rates between operator sectors.

Beyond the examples studied here in the context of fast and slow relaxation dynamics, the emergent graph geometry from quantum hitting times provides a new perspective on quantum dynamics and relaxation in terms of the geometric structure of the many-body state graph. 
In this context, a natural open question is whether other dynamical phenomena, including different forms of breakdown of thermalization, can also be characterized within this framework.
This suggests that quantum hitting times may provides a framework for exploring dynamical regimes of quantum many-body systems through emergent graph geometry.

\begin{acknowledgments}
\paragraph{Acknowledgments}
The authors thank Markus Heyl, Fabian Heidrich-Meisner, Roderich Moessner, and Monika Aidelsburger for valuable discussions and comments. This work was funded by the Deutsche Forschungsgemeinschaft (DFG, German Research Foundation) -- 499180199, 436382789, and 493420525; via FOR 5522 and large-equipment grants (GOEGrid cluster).
The authors gratefully acknowledge the resources on the LiCCA HPC cluster of the University of Augsburg, co-funded by the Deutsche Forschungsgemeinschaft (DFG, German Research Foundation) – Project-ID 499211671.

Research data associated with this article is available on Zenodo~\cite{this_zenodo}.

\end{acknowledgments}

\bibliography{main.bib}

@article{vanhorssen2015dynamics,
  author = {van Horssen, Merlijn and Levi, Emanuele and Garrahan, Juan P.},
  doi = {10.1103/physrevb.92.100305},
  issue = {10},
  journal = {Phys. Rev. B},
  month = {9},
  publisher = {American Physical Society (APS)},
  title = {Dynamics of many-body localization in a translation-invariant quantum glass model},
  url = {http://dx.doi.org/10.1103/physrevb.92.100305},
  volume = {92},
  year = {2015},
  pages = {100305}
}

@article{brighi2023hilbert,
  title={{H}ilbert space fragmentation and slow dynamics in particle-conserving quantum {E}ast models},
  author={Brighi, Pietro and Ljubotina, Marko and Serbyn, Maksym},
  journal={SciPost Phys.},
  volume={15},
  number={3},
  pages={093},
  year={2023},
  doi={10.21468/SciPostPhys.15.3.093}
}

@article{srednicki1994chaos,
  title={Chaos and quantum thermalization},
  author={Srednicki, Mark},
  journal={Phys. Rev. E},
  volume={50},
  number={2},
  pages={888},
  year={1994},
  publisher={APS},
  doi={10.1103/PhysRevE.50.888}
}

@article{deutsch2018eigenstate,
  title={Eigenstate thermalization hypothesis},
  author={Deutsch, Joshua M},
  journal={Rep. Prog. Phys.},
  volume={81},
  number={8},
  pages={082001},
  year={2018},
  publisher={IOP Publishing},
  doi={10.1088/1361-6633/aac9f1}
}

@misc{rakovszky2024bottlenecks,
  title={Bottlenecks in quantum channels and finite temperature phases of matter},
  author={Rakovszky, Tibor and Placke, Benedikt and Breuckmann, Nikolas P and Khemani, Vedika},
  archiveprefix = {arXiv},
  eprint = {2412.09598},
  year={2024}
}

@misc{karch2025probing,
  title={Probing quantum many-body dynamics using subsystem {L}oschmidt echos},
  author={Karch, Simon and Bandyopadhyay, Souvik and Sun, Zheng-Hang and Impertro, Alexander and Huh, SeungJung and Rodr{\'\i}guez, Irene Prieto and Wienand, Julian F and Ketterle, Wolfgang and Heyl, Markus and Polkovnikov, Anatoli and others},
  year={2025},
  archive={arXiv},
  eprint={arXiv:2501.16995}
}

@article{basko2006metal,
  title={Metal--insulator transition in a weakly interacting many-electron system with localized single-particle states},
  author={Basko, Denis M and Aleiner, Igor L and Altshuler, Boris L},
  journal={Ann. Phys.},
  volume={321},
  number={5},
  pages={1126--1205},
  year={2006},
  publisher={Elsevier},
  doi={10.1016/j.aop.2005.11.014}
}

@misc{ben2025many,
  title={Many-body cages: {D}isorder-free glassiness from flat bands in {F}ock space, and many-body {R}abi oscillations},
  author={Ben-Ami, Tom and Heyl, Markus and Moessner, Roderich},
  archiveprefix={arXiv},
  eprint={2504.13086},
  year={2025}
}

@misc{tan2025interference,
  title={Interference-caged quantum many-body scars: {T}he {F}ock space topological localization and interference zeros},
  author={Tan, Tao-Lin and Huang, Yi-Ping},
  archiveprefix={arXiv},
  eprint={2504.07780},
  year={2025}
}

@article{nicolau2025fragmentation,
  author = {Nicolau,  Eloi and Ljubotina,  Marko and Serbyn,  Maksym},
  title = {Fragmentation, Zero Modes, and Collective Bound States in Constrained Models},
  volume = {7},
  doi = {10.1103/sl79-1xgb},
  number = {1},
  journal = {PRX Quantum},
  year = {2026},
  pages = {010352},
  month = {Mar} 
}

@article{jonay2025localized,
  title = {Localized {F}ock space cages in kinetically constrained models},
  volume = {113},
  ISSN = {2469-9969},
  DOI = {10.1103/wz33-vczt},
  number = {13},
  journal = {Phys. Rev. B},
  publisher = {American Physical Society (APS)},
  author = {Jonay,  Cheryne and Pollmann,  Frank},
  year = {2026},
  month = {Apr},
  pages = {134313}
}

@article{venegas2012quantumw,
  title = {Quantum walks: {A} comprehensive review},
  volume = {11},
  ISSN = {1573-1332},
  url = {http://dx.doi.org/10.1007/s11128-012-0432-5},
  DOI = {10.1007/s11128-012-0432-5},
  number = {5},
  journal = {Quantum Inf. Process.},
  publisher = {Springer Science and Business Media LLC},
  author = {Venegas-Andraca,  Salvador Elías},
  year = {2012},
  month = jul,
  pages = {1015–1106}
}

@article{lan2018quantum,
  title = {Quantum Slow Relaxation and Metastability due to Dynamical Constraints},
  author = {Lan, Zhihao and van Horssen, Merlijn and Powell, Stephen and Garrahan, Juan P.},
  journal = {Phys. Rev. Lett.},
  volume = {121},
  issue = {4},
  pages = {040603},
  numpages = {6},
  year = {2018},
  month = {Jul},
  publisher = {American Physical Society},
  doi = {10.1103/PhysRevLett.121.040603},
  url = {https://link.aps.org/doi/10.1103/PhysRevLett.121.040603}
}

@article{coifman2006diffusionmaps,
    title = {Diffusion maps},
    journal = {Applied and Computational Harmonic Analysis},
    volume = {21},
    number = {1},
    pages = {5-30},
    year = {2006},
    issn = {1063-5203},
    doi = {https://doi.org/10.1016/j.acha.2006.04.006},
    url = {https://www.sciencedirect.com/science/article/pii/S1063520306000546},
    author = {Ronald R. Coifman and Stéphane Lafon},
}

@article{abanin2019colloquium,
  title={Colloquium: {M}any-body localization, thermalization, and entanglement},
  author={Abanin, Dmitry A and Altman, Ehud and Bloch, Immanuel and Serbyn, Maksym},
  journal={Rev. Mod. Phys.},
  volume={91},
  number={2},
  pages={021001},
  year={2019},
  publisher={APS},
  doi={10.1103/RevModPhys.91.021001}
}

@article{dallesio2016quantum,
  title={From quantum chaos and eigenstate thermalization to statistical mechanics and thermodynamics},
  author={D'Alessio, Luca and Kafri, Yariv and Polkovnikov, Anatoli and Rigol, Marcos},
  journal={Adv. Phys.},
  volume={65},
  number={3},
  pages={239--362},
  year={2016},
  publisher={Taylor \& Francis},
  doi={10.1080/00018732.2016.1198134}
}

@article{deutsch1991quantum,
  title={Quantum statistical mechanics in a closed system},
  author={Deutsch, Josh M},
  journal={Phys. Rev. A},
  volume={43},
  number={4},
  pages={2046},
  year={1991},
  publisher={APS},
  doi={10.1103/PhysRevA.43.2046}
}

@article{royen2024enhanced,
  title = {Enhanced many-body localization in a kinetically constrained model},
  author = {Royen, Karl and Mondal, Suman and Pollmann, Frank and Heidrich-Meisner, Fabian},
  journal = {Phys. Rev. E},
  volume = {109},
  issue = {2},
  pages = {024136},
  numpages = {9},
  year = {2024},
  month = {Feb},
  publisher = {American Physical Society},
  doi = {10.1103/PhysRevE.109.024136},
  url = {https://link.aps.org/doi/10.1103/PhysRevE.109.024136}
}

@misc{SM,
	note = {See Supplemental Material for further details on the quantum {E}ast and Triangular Lattice Gas model {H}amiltonians and details on the analytical calculations and proofs for the hitting matrix.}
}

@article{rigol2008thermalization,
  title = {Thermalization and its mechanism for generic isolated quantum systems},
  volume = {452},
  ISSN = {1476-4687},
  url = {http://dx.doi.org/10.1038/nature06838},
  DOI = {10.1038/nature06838},
  number = {7189},
  journal = {Nature},
  publisher = {Springer Science and Business Media LLC},
  author = {Rigol,  Marcos and Dunjko,  Vanja and Olshanii,  Maxim},
  year = {2008},
  month = Apr,
  pages = {854–858}
}

@article{kempe2005discrete,
  title = {Discrete Quantum Walks Hit Exponentially Faster},
  volume = {133},
  ISSN = {1432-2064},
  url = {http://dx.doi.org/10.1007/s00440-004-0423-2},
  DOI = {10.1007/s00440-004-0423-2},
  number = {2},
  journal = {Probab. Theory Relat. Fields},
  publisher = {Springer Science and Business Media LLC},
  author = {Kempe,  Julia},
  year = {2005},
  month = Feb,
  pages = {215–235}
}

@article{farhi1998quantum,
  title = {Quantum computation and decision trees},
  author = {Farhi, Edward and Gutmann, Sam},
  journal = {Phys. Rev. A},
  volume = {58},
  issue = {2},
  pages = {915--928},
  numpages = {0},
  year = {1998},
  month = {Aug},
  publisher = {American Physical Society},
  doi = {10.1103/PhysRevA.58.915},
  url = {https://link.aps.org/doi/10.1103/PhysRevA.58.915}
}

@article{menzler2025graph,
  title = {Graph theory and tunable slow dynamics in quantum {E}ast {H}amiltonians},
  author = {Menzler, Heiko Georg and Ba\~nuls, Mari Carmen and Heidrich-Meisner, Fabian},
  journal = {Phys. Rev. B},
  volume = {112},
  issue = {11},
  pages = {115141},
  numpages = {18},
  year = {2025},
  month = {Sep},
  publisher = {American Physical Society},
  doi = {10.1103/j7jf-746f},
  url = {https://link.aps.org/doi/10.1103/j7jf-746f}
}

@misc{menzler2026graph,
  archiveprefix = {arXiv},
  eprint = {2605.00094},
  author = {Menzler,  Heiko Georg and Świętek,  Rafał and Bañuls,  Mari Carmen and Heidrich-Meisner,  Fabian},
  title = {Graph-theory measures capture weak ergodicity breaking on large quantum systems},
  year = {2026},
}

@article{mlken2011continuous,
  title = {Continuous-time quantum walks: {M}odels for coherent transport on complex networks},
  volume = {502},
  ISSN = {0370-1573},
  url = {http://dx.doi.org/10.1016/j.physrep.2011.01.002},
  DOI = {10.1016/j.physrep.2011.01.002},
  number = {2-3},
  journal = {Physics Reports},
  publisher = {Elsevier BV},
  author = {M\"{u}lken,  Oliver and Blumen,  Alexander},
  year = {2011},
  month = May,
  pages = {37–87}
}

@article{rosenzweig1960repulsion,
	author = {Rosenzweig, Norbert and Porter, Charles E.},
	copyright = {http://link.aps.org/licenses/aps-default-license},
	doi = {10.1103/PhysRev.120.1698},
	issn = {0031-899X},
	journal = {Phys. Rev.},
	langid = {english},
	month = dec,
	number = {5},
	pages = {1698--1714},
	title = {\enquote{{R}epulsion of energy levels} in Complex Atomic Spectra},
	urldate = {2025-04-24},
	volume = {120},
	year = {1960}
}

@article{kravtsov2015random,
	author = {Kravtsov, V E and Khaymovich, I M and Cuevas, E and Amini, M},
	doi = {10.1088/1367-2630/17/12/122002},
	issn = {1367-2630},
	journal = {New J. Phys.},
	month = dec,
	number = {12},
	pages = {122002},
	title = {A Random Matrix Model with Localization and Ergodic Transitions},
	urldate = {2025-04-24},
	volume = {17},
	year = {2015}
}

@article{sala2020ergodicity,
  title={Ergodicity breaking arising from {H}ilbert space fragmentation in dipole-conserving {H}amiltonians},
  author={Sala, Pablo and Rakovszky, Tibor and Verresen, Ruben and Knap, Michael and Pollmann, Frank},
  journal={Phys. Rev. X},
  volume={10},
  number={1},
  pages={011047},
  year={2020},
  publisher={APS},
  doi={10.1103/PhysRevX.10.011047}
}

@article{yang2020hilbert,
  title={{H}ilbert-space fragmentation from strict confinement},
  author={Yang, Zhi-Cheng and Liu, Fangli and Gorshkov, Alexey V and Iadecola, Thomas},
  journal={Phys. Rev. Lett.},
  volume={124},
  number={20},
  pages={207602},
  year={2020},
  publisher={APS},
  doi={10.1103/PhysRevLett.124.207602}
}

@article{pancotti2020quantum,
  title = {Quantum {E}ast Model: {L}ocalization, Nonthermal Eigenstates, and Slow Dynamics},
  author = {Pancotti, Nicola and Giudice, Giacomo and Cirac, J. Ignacio and Garrahan, Juan P. and Ba\~nuls, Mari Carmen},
  journal = {Phys. Rev. X},
  volume = {10},
  issue = {2},
  pages = {021051},
  numpages = {21},
  year = {2020},
  month = {Jun},
  publisher = {American Physical Society},
  doi = {10.1103/PhysRevX.10.021051},
  url = {https://link.aps.org/doi/10.1103/PhysRevX.10.021051}
}

@article{coutinho2024selected,
  title = {Selected open problems in continuous-time quantum walks},
  volume = {12},
  ISSN = {2300-7451},
  url = {http://dx.doi.org/10.1515/spma-2024-0025},
  DOI = {10.1515/spma-2024-0025},
  number = {1},
  journal = {Special Matrices},
  publisher = {Walter de Gruyter GmbH},
  author = {Coutinho,  Gabriel and Guo,  Krystal},
  year = {2024},
  pages = {20240025},
  month = {Jan} 
}

@article{magniez2012hitting,
  title={On the hitting times of quantum versus random walks},
  author={Magniez, Fr{\'e}d{\'e}ric and Nayak, Ashwin and Richter, Peter C and Santha, Miklos},
  journal={Algorithmica},
  volume={63},
  number={1},
  pages={91--116},
  year={2012},
  publisher={Springer},
  doi={10.1007/s00453-011-9521-6}
}

@inproceedings{szegedy2004quantum,
  title={Quantum speed-up of {M}arkov chain based algorithms},
  author={Szegedy, Mario},
  booktitle={45th Annual IEEE symposium on foundations of computer science},
  pages={32--41},
  year={2004},
  organization={IEEE},
  doi={10.1109/FOCS.2004.53}
}

@dataset{this_zenodo,
    title={Identifying slow relaxation in many-body quantum systems through state-graph geometry and state-graph heterogeneity},
    author={Menzler, Heiko Georg and Ben-Ami, Tom},
    doi = {10.5281/zenodo.20547595},
    month={Jun},
    year={2026}
}

@book{meyer2000matrix,
  title = {Matrix Analysis and Applied Linear Algebra},
  ISBN = {9780898714548},
  DOI = {10.1137/1.9780898719512},
  publisher = {SIAM},
  author = {Meyer,  Carl},
  year = {2000},
  month = {Jan} 
}

@inbook{moore2002quantumwalks,
  title = {Quantum Walks on the Hypercube},
  ISBN = {9783540457268},
  ISSN = {0302-9743},
  url = {http://dx.doi.org/10.1007/3-540-45726-7_14},
  DOI = {10.1007/3-540-45726-7_14},
  booktitle = {Randomization and Approximation Techniques in Computer Science},
  publisher = {Springer Berlin Heidelberg},
  author = {Moore,  Cristopher and Russell,  Alexander},
  year = {2002},
  pages = {164–178}
}

@article{shenvi2003quantumrandom,
  title = {Quantum random-walk search algorithm},
  volume = {67},
  ISSN = {1094-1622},
  DOI = {10.1103/physreva.67.052307},
  number = {5},
  journal = {Phys. Rev. A},
  publisher = {American Physical Society (APS)},
  author = {Shenvi,  Neil and Kempe,  Julia and Whaley,  K. Birgitta},
  year = {2003},
  month = {May},
  pages = {052307}
}

@article{childs2002anexample,
  title = {An Example of the Difference Between Quantum and Classical Random Walks},
  volume = {1},
  ISSN = {1573-1332},
  DOI = {10.1023/a:1019609420309},
  number = {1-2},
  journal = {Quantum Inf. Process.},
  publisher = {Springer Science and Business Media LLC},
  author = {Childs,  Andrew M. and Farhi,  Edward and Gutmann,  Sam},
  year = {2002},
  month = Apr,
  pages = {35–43}
}

@article{childs2004spatial,
  author = {Childs,  Andrew M. and Goldstone,  Jeffrey},
  title = {Spatial search by quantum walk},
  volume = {70},
  doi = {10.1103/physreva.70.022314},
  journal = {Phys. Rev. A},
  year = {2004},
  month = {Aug},
  pages = {022314}
}

@misc{aldous2014fill,
    author = {Aldous, David and Fill, James Allen},
    title = {Reversible {M}arkov Chains and Random Walks on Graphs},
    year = {2002},
    note = {Unfinished monograph, recompiled 2014, available 
    at \url{www.stat.berkeley.edu/~aldous/RWG/book.html}}
}

@article{chandra1996theelectrical,
  title = {The electrical resistance of a graph captures its commute and cover times},
  volume = {6},
  ISSN = {1420-8954},
  DOI = {10.1007/bf01270385},
  number = {4},
  journal = {Comput. Complexity},
  publisher = {Springer Science and Business Media LLC},
  author = {Chandra,  Ashok K. and Raghavan,  Prabhakar and Ruzzo,  Walter L. and Smolensky,  Roman and Tiwari,  Prasoon},
  year = {1996},
  month = Dec,
  pages = {312–340}
}

@inproceedings{lovasz1993random,
  title={Random walks on graphs: {A} survey},
  author={Lov{\'a}sz, L{\'a}szl{\'o}},
  booktitle={Combinatorics, Paul Erd{\"o}s is eighty},
  volume={2},
  pages={1-46},
  year={1993}
}

@book{doyle1984random,
  title={Random walks and electric networks},
  author={Doyle, Peter G and Snell, J Laurie},
  volume={22},
  year={1984},
  publisher={American Mathematical Soc.}
}

@book{redner2001guide,
  title={A guide to first-passage processes},
  author={Redner, Sidney},
  year={2001},
  publisher={Cambridge university press}
}

@article{klein1993resistance,
  title = {Resistance distance},
  volume = {12},
  ISSN = {1572-8897},
  DOI = {10.1007/bf01164627},
  number = {1},
  journal={J. Math. Chem.},
  publisher = {Springer Science and Business Media LLC},
  author = {Klein,  D. J. and Randić,  M.},
  year = {1993},
  month = {Dec},
  pages = {81–95}
}

@inproceedings{aharonov2001quantumwalks,
  series = {STOC01},
  title = {Quantum walks on graphs},
  DOI = {10.1145/380752.380758},
  booktitle = {Proceedings of the thirty-third annual ACM symposium on Theory of computing},
  publisher = {ACM},
  author = {Aharonov,  Dorit and Ambainis,  Andris and Kempe,  Julia and Vazirani,  Umesh},
  year = {2001},
  month = {Jul},
  pages = {50–59},
  collection = {STOC01}
}

\clearpage

\appendix
\onecolumngrid
\begin{center}
{\large \bf End matter}\\
\end{center}
\twocolumngrid

\paragraph{Discrete-time evolution}
In many practical cases, the whole time-dependence of the kernel function $k_{ij}(T)$ is not available which constrains the ability of obtaining the value of the integral within $k_{ij}(T)$.
One particularly important situation where this is the case is in experimental setups.
In such cases, the integral in the definition of $k_{ij}$ has to be estimated similarly to numerical integration by discretization of the integrand in time steps $\Delta t$.
In the language of graph theory, the continuous quantum walk has to be estimated by a discrete quantum walk.
In addition to its importance to experimental measurements of $k_{ij}$, the discrete version of the quantum walk is itself an interesting object of study as we can use it to also study quantum systems that implement evolution under a discrete unitary $\qmoperator U$ like Floquet-driven systems or unitary circuits.
The corresponding discrete kernel function for $N$ steps is 
\begin{align}
    k_{ij}(N) = \sum_{n=0}^N |\mel{i}{\qmoperator U^n}{j}|^2\,.
    \label{eq:def_kernel_discrete}
\end{align}
Here we write $k_{ij}(N)$ to distinguish the discrete kernel function from the continuous case $k_{ij}(T)$.
For the continuous-time system the unitary $\qmoperator U = e^{-i \Delta t \qmoperator H}$ implements the discrete estimation of the evolution up to a time $T=N\Delta t$.
Notice, to connect to the discretized integral of $k_{ij}(T)$ one would include a factor $\Delta t$ on the right-hand-side of \cref{eq:def_kernel_discrete}.
For a more general setting, $\qmoperator U$ can be an arbitrary unitary implementing, e.g., a random unitary circuit.
Notice, that in a more general case, e.g., for random unitary circuits, $\qmoperator U$ may vary on each layer of the circuit.
To include this case in our formalism we may replace $\qmoperator U^n \to \prod_{i=1}^n \qmoperator U_n$ in \cref{eq:def_kernel_discrete}, where $\qmoperator U_n$ corresponds to the application of the circuit at layer~$n$.

In the case of a fixed $\qmoperator U_n = \qmoperator U$, we may find~---~as for the continuous case~---~a simplified expression of \cref{eq:def_kernel_discrete} by expanding in the eigenbasis that diagonalizes $\qmoperator U$,
\begin{align}
    k_{ij}(N) 
    &= \sum_{n=0}^N \sum_{\nu, \mu} e^{-i \omega_{\nu \mu} n} \braket{i}{\nu} \braket{\nu}{j}\braket{j}{\mu} \braket{\mu}{i}\nonumber\\ 
    &= \sum_{\nu \le \mu} \braket{i}{\nu} \braket{\nu}{j}\braket{j}{\mu} \braket{\mu}{i} \sum_{n=0}^N \cos(\omega_{\nu \mu}n)\,,
\end{align}
where we used the short-form $\omega_{\nu\mu} = \phi_\nu -\phi_\mu$, with $e^{i\phi_{\nu/\mu}}$ being the eigenvalues of $\qmoperator U$, such that for the case of a discrete evolution of a continuous-time Hamiltonian  dynamics $\phi_\nu = E_\nu \Delta t$. 
Next we write $g(N) = \sum_{n=1}^N \cos(\omega_{\nu\mu} n) = \sin(N\omega_{\nu\mu}/2)\cos((N+1)\omega_{\nu\mu}/2)/\sin(\omega_{\nu\mu}/2)$.
This already simplifies the calculation of $k_{ij}(N)$ immensely, allowing us to calculate $k_{ij}(N)$ without summing over previous terms.
We can also analyze $\lim_{N\to\infty} g(N)$ which is, in general, oscillating but does permit a sensible limit for large $N$.
At large $N$, $g(N)$ is dominated by the contribution of $\omega_{\nu \mu} = 0$, which are the only terms not oscillating with $N$.
We can write 
\begin{align}
  g(N) = \begin{cases}
      \omega_{\nu\mu} = 0, \quad N\\
      \omega_{\nu\mu} \neq 0,  \quad \mathcal{O}(1)
  \end{cases}\,,  
\end{align}
which can be obtained from the limit
$\lim_{\omega_{\nu\mu} \to 0} g(N) =\lim_{\omega_{\nu\mu} \to 0}\sin(N\omega_{\nu\mu}/2)/\sin(\omega_{\nu\mu}/2) = N$ and from upper-bounding the periodic behavior of $g(N)$ when $\omega_{\nu\mu} \neq 0$ with a constant value.
Therefore, at large $N$ we can ignore the oscillatory terms, obtaining
\begin{align}
    \lim_{N\to\infty} k_{ij}(N) / N = \sum_{\nu} |\braket{i}{\nu}|^2 |\braket{j}{\nu}|^2\,.
\end{align}
This is the same expression as we have previously obtained for the late-time limit in the continuous case because when we insert $N = T/\Delta t$ and $\phi_\nu = E_\nu\Delta t$ the expression becomes that of \cref{eq:kernel_continuous_limit} up to a factor of $\Delta t$ that was omitted in \cref{eq:def_kernel_discrete}.
The calculations are explained in greater detail in~\cite{SM}.

This shows, that the discrete kernel function at late times carries the same information as the continuous one and therefore, when comparing discrete and the continuous evolution of the same system, the discrete kernel function may deviate from the continuous kernel function at small $N$, however it will not distinguish between the continuous and discrete systems at late times.

\paragraph{Extracting positions of crossing points}

\begin{figure}
    \centering
    \begin{overpic}{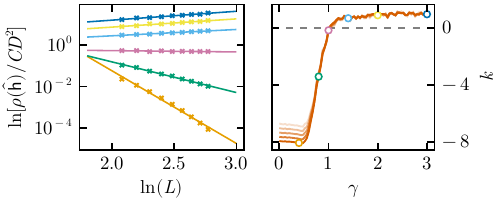}
    \put(19, 14){(a)}
    \put(56, 29){(b)}
    \put(75, 14){RPM}
    \end{overpic}
    \begin{overpic}{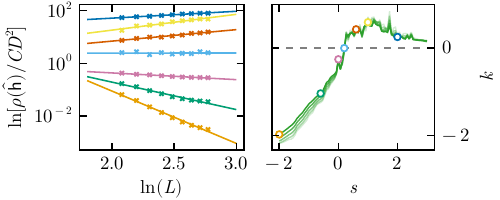}
    \put(19, 14){(c)}
    \put(56, 34){(d)}
    \put(75.5, 14){QEM}
    \end{overpic}
    \begin{overpic}{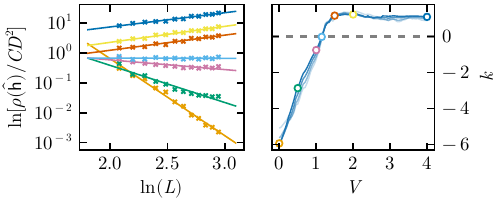}
    \put(19, 14){(e)}
    \put(56, 34.5){(f)}
    \put(73, 14){TLGM}
    \end{overpic}
    {\phantomsubcaption\label{fig:scaling_analysis_rpm_fit}}
    {\phantomsubcaption\label{fig:scaling_analysis_rpm_cross}}
    {\phantomsubcaption\label{fig:scaling_analysis_qe_fit}}
    {\phantomsubcaption\label{fig:scaling_analysis_qe_cross}}
    {\phantomsubcaption\label{fig:scaling_analysis_tlg_fit}}
    {\phantomsubcaption\label{fig:scaling_analysis_tlg_cross}}
    \caption{
        Finite-size scaling analysis of $\rho(\qmoperator{\mathsf{h}})$ for the RPM [\subref{fig:scaling_analysis_rpm_fit}-\subref{fig:scaling_analysis_rpm_cross}],
        QEM [\subref{fig:scaling_analysis_qe_fit}-\subref{fig:scaling_analysis_qe_cross}],
        and TLGM [\subref{fig:scaling_analysis_tlg_fit}-\subref{fig:scaling_analysis_tlg_cross}]. The crossing point is determined from the finite-size scaling exponent.
        The exponent $k$ is obtained from a linear fit in log-log space, as $\ln[\rho(\qmoperator{\mathsf{h}})] \propto L^k$.
        The left column [\subref{fig:scaling_analysis_rpm_fit}, \subref{fig:scaling_analysis_qe_fit}, \subref{fig:scaling_analysis_tlg_fit}] shows the log-log plots of $\ln[\rho(\qmoperator{\mathsf{h}})]$ versus $L$ together with the linear fits.
        The right column [\subref{fig:scaling_analysis_rpm_cross}, \subref{fig:scaling_analysis_qe_cross}, \subref{fig:scaling_analysis_tlg_cross}] shows $k$ as a function of the tuning parameter. 
        The dashed vertical line corresponds to $k=0$ where $\ln[\rho(\qmoperator{\mathsf{h}})]$ shows a flat scaling.
        Each marker color corresponds to a single tuning parameter value; the same color is used in the right column to show the finite-size fits. 
        To assess the finite-size dependence, we also show linear fits with the respective largest system sizes removed in lighter colors in the right column.
    }
    \label{fig:scaling_analysis}
\end{figure}

The change in the structure of the hitting matrix, quantified by $\rho(\qmoperator{\mathsf{h}})$, provides a highly sensitive quantitative tool for extracting the point where slow dynamics suddenly appear.
As there is no established rigorous scaling theory, we adopt a finite-size scaling analysis the extrapolate the crossing point. This is limited by the small system sizes that are accessible to exact diagonalization, which can limit the accuracy of the extrapolation to the thermodynamic limit. 
\cref{fig:scaling_analysis} shows the results of our scaling analysis and the extraction of the crossing points of the three models.
The left column of \cref{fig:scaling_analysis} shows finite-size scaling of $\ln[\rho(\qmoperator{\mathsf{h}})]$ on a log-log plot, corresponding to the tuning parameter values as indicated by the values marked in the right column. We fit
\[
\ln[\rho(\qmoperator{\mathsf{h}})] = a L^k,
\]
where $a$ and $k$ are the fitting parameters.
In the right column, we show the fitted exponent $k$ as a function of system parameter.
The crossing point corresponds to when $k$ crosses from negative into positive values, indicating the breakdown of homogeneity.
To assess the accuracy of the finite system-size scaling ansatz, we additionally show the result for $k$ when excluding the last few system sizes from the fit in brighter colors.
This highlights where the naive scaling theory suffers from significant corrections.
In the RPM, the crossing point at the ergodic to fractal transition can be reliably extracted, reproducing the known transition point at $\gamma = 1$.
Only for $\gamma < 0.5$ do we find that the chosen fitting ansatz is inaccurate.
However, we observe, that when system sizes increase, the fitted $k$ increases, indicating that $\ln[\rho(\qmoperator{\mathsf{h}})]$ is decreasing even faster than polynomially.
The extracted crossing point is between $\gamma = 1.00$ and $\gamma = 1.04$, due to the discrete grid of the tuning parameter steps increasing by $0.04$ and consistent with the prediction of $\gamma=1$.
In contrast, the QEM and TLGM suffer from strong finite size effects, which severely limit the confidence in the naive interpolation method.
In particular, away from the crossing point, both models are subject to finite size corrections in the naive scaling ansatz.
For both cases, in the regimes of negative $k$, we observe that for larger system sizes, the data points systematically deviate from the linear fit, indicating that the naive ansatz is incomplete.
However, at the crossing point, both models show much smaller deviation when system size is increased and there is no obvious systematic error that violates the scaling theory.
We cannot exclude additional systematic effects for larger system sizes.
These limitations arise due to the accessible system sizes, and not due to the method based on the geometric characterization of the quantum dynamic.
The extrapolated crossing points are around $s\approx 0.2$ for the QEM and $V/\unit{\hopping} \approx 1.05$ for the TLGM.

\onecolumngrid

\pagebreak

\appendix
\section*{Supplementary Material: \\ Identifying slow relaxation in many-body quantum systems through\\state-graph geometry and state-graph heterogeneity}

\author{Heiko Georg Menzler}
\affiliation{Institut für Theoretische Physik, Georg-August-Universität Göttingen, D-37077 Göttingen, Germany}

\author{Tom Ben-Ami}
\affiliation{Theoretical Physics III, Center for Electronic Correlations and Magnetism, Institute of Physics, University of Augsburg, D-86135 Augsburg, Germany}
\affiliation{Max-Planck-Institut für Physik komplexer Systeme, Nöthnitzer Straße 38, Dresden 01187, Germany}

\maketitle

\setcounter{figure}{0}
\setcounter{equation}{0}
\setcounter{table}{0}
\setcounter{section}{0}

\renewcommand{\thetable}{S\arabic{table}}
\renewcommand{\thefigure}{S\arabic{figure}}
\renewcommand{\theequation}{S\arabic{equation}}
\renewcommand{\thepage}{S\arabic{page}}

\renewcommand{\thesection}{S\arabic{section}}

\section{Additional details on the models}

\subsection{Quantum East Model}

The quantum East model (QEM) is a paradigmatic model for slow dynamics induced by kinetic constraints~\cite{vanhorssen2015dynamics}.
The quantum East model realizes on of the most simple constraints, a one-spin-faciliated constraint, meaning that whether a transition is allowed or not allowed depends only on a single spin in a given configuration.
The Hamiltonian of the QEM is
\begin{align}
    \qmoperator H_\mathrm{QEM} = -\frac{1}{2}\sum_{\ell=1}^{L-1} e^{-s} \qmoperator n_\ell \qmoperator \sigma^x_{\ell+1} + \frac{1}{2}\sum_{\ell=1}^{L-1} \qmoperator n_\ell\,,
\end{align}
When the QEM is simulated with open boundary conditions it attains a block diagonal structure due to its constraint.
The block structure of the QEM Hamiltonian arises due to strings of unoccupied sites starting on the first site.
These strings are non-dynamical, because $\qmoperator H$ transitions where a site may become occupied can only occur when the previous site on the lattice is already occupied.
This means that the number of unoccupied sites starting from the first site is a good quantum number and leads to disconnected blocks of the Hamiltonian.

In our setting, this is however something that we would like to avoid and therefore we impose a boundary condition to the QEM where the site before the first site is actually occupied and facilitates dynamics. 
This site is non-dynamical and therefore, in effect, we simply add a term to the Hamiltonian of $-e^{-s} \qmoperator \sigma^x_1/2$.
Furthermore, there is another relevant symmetry which is generated by $\qmoperator\sigma_L^x$ and therefore we also allow the system to act on a non-dynamical site after the last site which is in an eigenstate of $\qmoperator \sigma^x$.
This term, induces a dephasing in the simulated system with a term $-e^{-s}\qmoperator n_L/2$.
Therefore, the Hamiltonian with boundary conditions as it is used throughout the manuscript is $\qmoperator H_\mathrm{QEM} - \frac{1}{2} e^{-s}(\qmoperator \sigma_1^x + n_L)$.
This ensures, that the QEM Hamiltonian does not host any symmetries.

\subsection{Triangular Lattice Gas Model}

In contrast to the QEM, the Triangular Lattice Gase model (TLGM) is an example of a more conventional glass model, where particles restrict each others movement. 
The constraint in the TLGM is exactly of such a form: particles can not move from one site to another if all joint neighbor sites are occupied. 
In its classical version, the TLGM is often simulated as a 2D model, however here we restrict the triangular lattice geometry to a ladder to allow simulation of large enough systems where we can see the influence of the constraint come into effect, see e.g., \cite{lan2018quantum,royen2024enhanced,menzler2026graph} for visualization of the lattice geometry.
The Hamiltonian that describes the model is
\begin{align}
    \qmoperator H_\mathrm{TLGM} 
    = 
    \sum_{\langle \ell,\ell^\prime \rangle} \qmoperator C_{\ell \ell^\prime} 
    \big[ 
    -\unit{\hopping}(\qmoperator a_\ell^\dagger \qmoperator a_{\ell^\prime} + \hc)
    + V (\qmoperator n_\ell (\mathbb{1} - \qmoperator n_{\ell^\prime}) + (\mathbb{1} - \qmoperator n_\ell) \qmoperator n_{\ell^\prime})
    \big]
    \,,
\end{align}
where $\qmoperator a^\dagger_\ell$ ($\qmoperator a_\ell$) are hard-core boson creation (annihilation) operators acting on site $\ell$.
The sum $\langle \ell, \ell^\prime\rangle$ runs over nearest-neighbor sites and the constraint $\qmoperator C_{\ell \ell^\prime} = \mathbb{1} - \prod_{k \in \mathcal{N}(\ell,\ell^\prime)} \qmoperator n_k$, with $\mathcal{N}(\ell,\ell^\prime)$ being the set of sites neighbor both to $\ell$ and $\ell^\prime$, restricts particles from hopping when all shared neighbor sites are occupied.
The Hamiltonian is particle conserving and as a conventional glass model, the TLGM is expected to show slow dynamics as particle density $N/L$ is increased but due to finite system restrictions a full understanding of particle density in the quantum system is so far elusive.
In the quantum case, the TLGM is also often studied when $V/\unit{\hopping}$ is varied, because large $V/\unit{\hopping}$ also can increase the strength of the constraint as configurations with high mobility are energetically penalized.
In \cite{lan2018quantum} it was shown, that when $V \gtrsim 1$, for certain initial states which have a high-particle-density subsystem, the decay of autocorrelation is metastable such that the autocorellations do decay, but they do so only at late times.

\section{Details on the calculations for hitting times}

Here we explicitly collect the details on the calculations for hitting time shown in the main manuscript.
The definition of the hitting time is 
\begin{align}
    \mathsf{H}_{ij}
    =
    \inf\{
        x: k_{ij}(x) \ge C
    \}\,.
    \label{eq:dist_def}
\end{align}
for some positive value of $C$.
The infimum in \cref{eq:dist_def} ensures that hitting time is well defined even in cases where $\ket{i}$ and $\ket{j}$ cannot be connected with respect to the unitary dynamics of $\qmoperator U$ in the chosen basis.
In such cases the hitting time is infinite.
In cases where the dynamics induced by $\qmoperator U$ are irreducible (all basis states are connected), we may replace the infimum with the minimum, as we do in the main text.

\subsection{Late-time limit}
First we show that for the kernel the hitting times function $k_{ij}$ has an asymptotic form $\lim_{x\to\infty} k_{ij}(x) \propto \alpha_{ij} x$, where $\alpha_{ij}$ is a simple proportionality factor.
For the case of the continuous kernel function $k_{ij}(T)$ from the main text, we find
\begin{align}
    k_{ij}(T) 
    &=
    \int_{0}^{T} | \mel{i}{\qmoperator U(t)}{j} | ^ 2 \,\mathrm{d}t
    =
    \frac{1}{2}\int_{-T}^{T} | \mel{i}{\qmoperator U(t)}{j} | ^ 2 \,\mathrm{d}t
    =
    \frac{1}{2}\int_{-T}^{T} \mel{i}{\qmoperator U(t)}{j} \mel{j}{\qmoperator U(t)^\dagger}{i} \,\mathrm{d}t
    \nonumber\\
    &=
    \frac{1}{2}\int_{-T}^{T}  \sum_{\nu, \mu}\braket{i}{\nu}\braket{\nu}{j} \braket{j}{\mu}\braket{\mu}{i}e^{-i (E_\nu - E_\mu) t}\,\mathrm{d}t
    =
    \sum_{\nu,\mu} \braket{i}{\nu}\braket{\nu}{j} \braket{j}{\mu}\braket{\mu}{i} \frac{\sin(\omega_{\nu\mu}T)}{\omega_{\nu\mu}}\,,
\end{align}
where in the first step we used $|\mel{i}{\qmoperator U(t)}{j}|^2 = \mel{i}{\qmoperator U(t)}{j}\mel{j}{\qmoperator U^\dagger(t)}{i} = \mel{i}{\qmoperator U^\dagger(-t)}{j}\mel{j}{\qmoperator U(-t)}{i} = |\mel{i}{\qmoperator U(-t)}{j}|^2$ and in the last step we defined $\omega_{\nu\mu} = E_\nu - E_{\mu}$ for the eigenpairs $E_\nu, \ket{\nu}$ of $\qmoperator H$ in $\qmoperator U(t) = e^{-it\qmoperator H}$. 
We can split the double sum over eigenstates into diagonal/degenerate and off-diagonal contributions
\begin{align}
    k_{ij}(T) = \sum_{\nu, \mu: E_\nu = E_\mu} \braket{i}{\nu}\braket{\nu}{j} \braket{j}{\mu}\braket{\mu}{i} T
    +
    \sum_{\nu, \mu :E_\nu\neq E_\mu}\braket{i}{\nu}\braket{\nu}{j} \braket{j}{\mu}\braket{\mu}{i} \frac{\sin(\omega_{\nu\mu}T)}{\omega_{\nu\mu}}\,,
\end{align}
where, we used $\lim_{\omega\to 0} \sin(\omega T)/\omega = T$.
Under the assumption that $T$ is large, the second, oscillatory term becomes negligible $\lim_{T\to\infty} k_{ij}(T)/T = \alpha_{ij}$ with $\alpha_{ij} = \sum_{E_\nu = E_\mu} \braket{i}{\nu}\braket{\nu}{j} \braket{j}{\mu}\braket{\mu}{i}$. 
Hence, it is clear that for large $T$ we find a linearly increasing behavior for the kernel function $k_{ij}$ in $T$.
Note, that in the main text we have assumed no degeneracies $E_\nu \neq E_{\mu}$, but here we provide the full expression.
Further, any type of degeneracies do not hurt the core argument for linear growth, as the non-oscillatory terms due to these degeneracies can always be absorbed into $\alpha_{ij}$.  

Next, we also go through the same argument for the discrete kernel function $k_{ij}(N)$ for the case that $\qmoperator U_n = \qmoperator U$: 

\begin{align}
    k_{ij}(N) 
    &= 
    \sum_{n=0}^N |\mel{i}{\qmoperator U^n}{j}|^2
    =
    \frac{1}{2}\sum_{n=-N}^N \mel{i}{\qmoperator U^n}{j}\mel{j}{\qmoperator U^{-n}}{i}
    =
    \frac{1}{2}\sum_{n=-N}^N \sum_{\nu,\mu} \braket{i}{\nu}\braket{\nu}{j}\braket{j}{\mu}\braket{\mu}{i} e^{-i n (\phi_\nu - \phi_\mu)}
    \nonumber\\
    &=
    \sum_{\nu,\mu} \braket{i}{\nu}\braket{\nu}{j}\braket{j}{\mu}\braket{\mu}{i}\sum_{n=0}^N  \cos(n \omega_{\nu\mu})
    = 
    \sum_{\nu,\mu} \braket{i}{\nu}\braket{\nu}{j}\braket{j}{\mu}\braket{\mu}{i} \frac{\sin((N+1)\omega_{\nu\mu}/2)\cos(N\omega_{\nu\mu}/2)}{\sin(\omega_{\nu\mu}/2)}
    \,,
\end{align}
where $\omega_{\nu\mu} = \phi_\nu-\phi_\mu$ with the eigenpairs $e^{i\phi_\nu}, \ket{\nu}$ that diagonalize $\qmoperator U$.
For the last step we used $\sum_{n=0}^N \cos(\omega n) = \mathrm{Re} \sum_{n=0}^N e^{i\omega n}$ and the geometric series $\sum_{n=0}^N e^{i\omega n} = \frac{1 - e^{i\omega(N+1)}}{1-e^{i\omega}} = e^{i\omega N/2} \frac{\sin((N+1)\omega / 2)}{\sin(\omega/2)}$ before again taking the real part.
As for the continuous case, we may decompose this expression into
\begin{align}
    k_{ij}(N) = \sum_{\nu,\mu:\phi_\nu=\phi_\mu} \braket{i}{\nu}\braket{\nu}{j}\braket{j}{\mu}\braket{\mu}{i} N + \sum_{\nu,\mu:\phi_\nu \neq \phi_\mu} \braket{i}{\nu}\braket{\nu}{j}\braket{j}{\mu}\braket{\mu}{i}\frac{\sin((N+1)\omega_{\nu\mu}/2)\cos(N\omega_{\nu\mu}/2)}{\sin(\omega_{\nu\mu}/2)}\,,
\end{align}
where we again find, that there is one term that scales linearly with $N$ with the prefactor $\alpha_{ij} = \sum_{\phi_\nu = \phi_\mu} \braket{i}{\nu}\braket{\nu}{j}\braket{j}{\mu}\braket{\mu}{i}$ which is similar to the one we found before in the continuous case and another term which oscillates.
Again we can write $\lim_{N\to\infty}k_{ij}(N)/N = \alpha_{ij}$.
As mentioned in the End matter of the main manuscript, this shows that when discretizing the continuous case using $\qmoperator U = e^{- i \Delta t \qmoperator H}$ and therefore $\phi_\nu = \Delta t E_\nu$, we recover the same result at late times.

\subsection{Relaxation of autocorrelations}

The relaxation of autocorrelation functions under the Hamiltonian $\qmoperator H$ for operators which are diagonal in the graph basis can be related to the distributions of $\mathsf{H}_{ij}$. 
We can show this directly by studying the infinite temperature autocorellation $c(t) = \frac{1}{D}\Tr(\qmoperator n(t) \qmoperator n)$ of some operator $\qmoperator n = \sum_{i \in S} \ketbra{i}{i}$ with $S = \{i : \qmoperator n \ket{i} = \ket{i}\}$, where $D$ is Hilbert space dimension.
First we simplify

\begin{align}
  \Tr(\qmoperator n(t)\qmoperator n) &= \sum_{i} \mel{i}{\qmoperator n(t)\qmoperator n}{i} 
  = 
  \sum_{i\in S} \mel{i}{\qmoperator n(t)}{i}
  =
  \sum_{i\in S} \mel{i}{e^{i t \qmoperator H} \qmoperator n e^{-i t \qmoperator H}}{i}
  =
  \sum_{i\in S} \mel{i}{e^{i t \qmoperator H} \left[\sum_{j \in S} \ketbra{j}{j}\right] e^{-i t \qmoperator H}}{i}
  \nonumber\\
  &= 
  \sum_{i\in S}\sum_{j \in S} \mel{i}{e^{i t \qmoperator H}}{j}\mel{j}{e^{-i t \qmoperator H}}{i}
  = 
  \sum_{(i, j)\in S^2} |\braket{i(t)}{j}|^2
  \,,
\end{align}
where we have used to short-form $S^2 = S\times S$.
This allows us to further simplify the time-averaged autocorrelation function $\overline{c}(T) = \frac{1}{T}\int_0^T c(t) \,\mathrm{d}t$
\begin{align}
    \overline{c}(T)
    =
    \frac{1}{T}\int_0^T\frac{1}{D}\Tr(\qmoperator n(t)\qmoperator n) \,\mathrm{d}t
    =
    \frac{1}{DT} \int_0^T \sum_{(i, j) \in S^2} |\braket{i(t)}{j}|^2\,\mathrm{d}t
    = 
    \frac{1}{DT}\sum_{(i, j)\in S^2} k_{ij}(T)\,.
\end{align}
Here, $k_{ij}$ is again the kernel function from before.
Now taking the large-time limit recovers exactly
\begin{align}
    \lim_{T\to\infty} \overline{c}(T)
    = 
    \lim_{T\to\infty} \frac{1}{DT} \sum_{(i, j)\in S^2} k_{ij}(T)
    =
    \frac{1}{D} \sum_{(i, j)\in S^2} \lim_{T\to\infty} k_{ij}(T)/T
    =
    \frac{1}{D} \sum_{(i, j)\in S^2} (\mathsf{H}_{ij}/C)^{-1}
    =
    \frac{C}{D} \sum_{(i, j)\in S^2} \mathsf{H}_{ij}^{-1}
    \,.
\end{align}

Now we want to consider the case where the autocorrelation $c_{\ell}(t) = \frac{1}{D}\Tr(\qmoperator n_\ell(t) \qmoperator n_\ell)$ is averaged over multiple projectors $\qmoperator n_\ell$, $\ell = 1,\ldots,L$ with their corresponding sets $S_\ell$.
For such a case we require that $\frac{1}{L}\sum_\ell \qmoperator n_\ell / s = \mathbb{1}$, where $s = \Tr(n_\ell)/D\,\forall\ell$.
In effect, this condition just means that across all sets $S_\ell$, every state occurs equally many times and that all sets $S_\ell$ have the same cardinality.
This is a case which is true, for example, in the case when $\qmoperator n_\ell$ are density operators on a spin-$1/2$ chain but it would also be true in the case when $\qmoperator n_\ell = \ketbra{\ell}{\ell}$.
For normalization we will use that $c_\ell(0) / s =1$.
We write the averaged autocorrelation $A(t) = \frac{1}{L}\sum_{\ell=1}^L s^{-1}c_\ell(t)$, for which $A(0) = 1$.
We obtain
\begin{align}
    \lim_{T\to\infty} \overline{A}(T)
    &=
    \frac{1}{Ls}\sum_{\ell=1}^L\lim_{T\to\infty}\overline c_\ell(T)
    =
    \frac{1}{Ls}\sum_{\ell=1}^L 
    \left[\frac{C}{D}\sum_{(i,j)\in S_\ell^2}
    \mathsf{H}_{ij}^{-1}\right]
    = 
    \frac{1}{Ls D^2}\sum_{\ell=1}^L 
    \sum_{(i,j)\in S_\ell^2}
    (\mathsf{H}_{ij}/CD)^{-1}
    \nonumber\\
    &= 
    \frac{1}{D^2}
    \sum_{i, j}
    (\mathsf{H}_{ij}/CD)^{-1}
    =
    CD\mathbb{E}[\mathsf{H}_{ij}^{-1}]
    \,,
\end{align}

where we use $\mathbb{E}[\cdot]$ to denote the average over all pairs $i, j$.
In the second to last step we have used, that when all basis states show up equally in $\{S_\ell\}$ and all $\{S_\ell\}$ have the same cardinality, then also all pairs of basis states $(i,j)\in S_\ell^2$ will show up a number of $sL$-times.
From this result, we can clearly see, how the distribution of $\mathsf{H}_{ij}/CD$, discussed in the main text, is related to the relaxation of autocorrelation functions.

The case described above, mirrors exactly the case of a density-density infinite-temperature autocorrelation function in a particle-conserving system which is averaged over all spatial sites.
Here the mean of the distribution of $\mathsf{H}_{ij}$ gives the late-time value of the autocorrelation function.
Notice however, that for this it was necessary to average over all $c_\ell(t)$.
It is clear, that when $\mathsf{H}_{ij}$ is sharply peaked, every $\overline{c}_\ell$ will also have the same mean value, as every sample of the distribution of $\mathsf{H}_{ij}$ is representative of its mean.
In the case of a broad distribution this is however not the case at all:
In a reasonable scenario, where $\mathsf{H}_{ij}$ is broad, we find that $\overline{A}$ decays to the mean of the distribution, while individual $\overline{c}_\ell$ may vary strongly --- depending on how the sets $S_\ell$ sample the distribution of $\mathsf{H}_{ij}$.
This makes the width of the distribution a sensitive probe for possible heterogeneity.
We can rephrase this result in a more precise way:
Whenever the distribution of $\mathsf{H}_{ij}$ is sharply peaked, every possible choice of projectors $\{\qmoperator n_\ell\}$, which are diagonal in the graph's basis, will show universally fast decay across all $c_\ell$.
Whenever the distribution of $\mathsf{H}_{ij}$ is broad, it is always possible to choose a set of $\{\qmoperator n_\ell\}$ such that $c_\ell$ are heterogeneously decaying.
Crucially, depending on how any operator $\qmoperator n_\ell$ samples the distribution of $\mathsf{H}_{ij}$, it can either look non-thermal or thermal.
Therefore, even observing single autocorrelation functions $c_\ell$ can obscure the detection of slow dynamics regimes and the distribution of $\mathsf{H}_{ij}$ gives a more fine-grained characterization of the heterogeneous decay on the systems state graph.

\section{\texorpdfstring{Proof that $\rho(\qmoperator{\mathsf{H}})$ is minimal for uniform $\qmoperator{\mathsf{H}}$}{Proof that ρ(h) is minimal for uniform h}}

We want to show that the spectral radius $\rho(\qmoperator{\mathsf{H}})$ is minimal with the corresponding eigenvector $x = (1, 1, 1, \ldots, 1)$, iff $\mathsf{H}_{ij} = \mathrm{const.}\,$.
We first prove the case which occurs in discrete walks where $\mathsf{H}_{ii} = 0$, but it turns out that this requirement does not impact the proof up to the value of the spectral radius $\rho(\qmoperator{\mathsf{H}})$.

\begin{theorem}
    Let $\qmoperator{\mathsf{H}}$ be a symmetric and positive but hollow matrix (${\mathsf{H}}_{ij} = \mathsf{H}_{ji} \ge c > 0\,\forall i \neq j$ with some constant $c$ and ${\mathsf{H}}_{ii} = 0$) with spectral radius $\rho(\qmoperator{\mathsf{H}})$ corresponding to the eigenvector $x$ and $\mathrm{dim}(\qmoperator{\mathsf{H}}) = N$.
    Then $\rho(\qmoperator{\mathsf{H}}) = c(N-1)$ is minimal and $\mathsf{H}_{ij} = c \, \forall i \neq j$ with 
    $x = (1, 1, 1, \ldots, 1)^T$.
\end{theorem}

\noindent{}Proof:
Let $x$ with $x_i>0$ be an eigenvector of $\qmoperator{a} = c(\qmoperator J - \mathbb{1})$ with $(\qmoperator J)_{ij} = 1$. 
As $x = (1, 1, 1, \ldots, 1)^T$ is an eigenvector of $\qmoperator a$, by the Perron-Frobenius theorem, $x$ is the eigenvector with the eigenvalue $\rho(\qmoperator a)$ for which $\qmoperator a x = \rho(\qmoperator a)x = c (N - 1) x$.
Now let also $\qmoperator b$ be a symmetric and positive but hollow matrix, with an eigenvector $y$ with $y_i > 0$, but we assume that least one entry of $\qmoperator b$ is larger than $\qmoperator a$, therefore $b_{ij} \ge a_{ij}$ but $\qmoperator b \neq \qmoperator a$.
We write $y^T(\qmoperator b - \qmoperator a) x = \sum_{ij} y_i (b_{ij} - a_{ij}) x_j > 0$ where we used that $y_i x_j > 0$ and $b_{ij} \ge a_{ij}$ with at least one relation being strict because $\qmoperator b \neq \qmoperator a$.
Therefore we can also write $ 0 < y^T(\qmoperator b - \qmoperator a) x = y^T \qmoperator b x - y^T \qmoperator a x = \rho(\qmoperator b) y^T x - \rho(\qmoperator a) y^T x \Leftrightarrow \rho(\qmoperator a) < \rho(\qmoperator b)$, where we used $y^T\qmoperator b = \rho(\qmoperator b)y^T$ and $\qmoperator a x = \rho(\qmoperator a) x$ and again $y^T x > 0$ due to $y_i x_i > 0$ .
This shows that choosing any $\qmoperator b$ with $\rho(\qmoperator b) \le \rho(\qmoperator a)$ and $\qmoperator b \neq \qmoperator a$ leads to a contradiction, implying that $\rho(\qmoperator a)$ is minimal only for $\qmoperator a = c(\qmoperator J - \mathbb{1})$, completing the proof.
\vspace{1em}

In the proof we have used a result from the Perron-Frobenius theorem (see for example~\cite{meyer2000matrix}) which states, that there is only a single eigenvector $x$ for which $x_i > 0$ and this is exactly the eigenvector which correspond to the largest eigenvalue $\rho(\qmoperator{\mathsf{H}})$.

\begin{theorem}
    Let $\qmoperator{\mathsf{H}}$ be a symmetric and positive matrix (${\mathsf{H}}_{ij} = \mathsf{H}_{ji} \ge c > 0$ with some constant $c$) with spectral radius $\rho(\qmoperator{\mathsf{H}})$ corresponding to the eigenvector $x$ and $\mathrm{dim}(\qmoperator{\mathsf{H}}) = N$.
    Then $\rho(\qmoperator{\mathsf{H}}) = cN$ is minimal and $\mathsf{H}_{ij} = c$ with 
    $x = (1, 1, 1, \ldots, 1)^T$.
\end{theorem}

Proof: Same as above up to $\qmoperator a = c \qmoperator J$.
$x$ is still an eigenvector of $\qmoperator a$ with $\qmoperator a x = c N x$ and the rest follows in the same way.
\vspace{1em}

Based on this we argue that $\qmoperator{\mathsf{H}}$ has a universal structure whenever $\rho(\qmoperator{\mathsf{H}})$ is minimal.



\end{document}